\documentclass[journal, 10pt ,twocolumn]{IEEEtran}
\usepackage{mathrsfs}
\usepackage{graphicx}
\usepackage{subfigure}
\usepackage{multirow}
\usepackage{amsfonts}
\usepackage{amssymb}
\usepackage{amsmath}
\usepackage{url}
\usepackage{lineno}
\usepackage{latexsym}
\usepackage{amsxtra}
\usepackage{array}
\usepackage{color}
\usepackage{subfigure}
\usepackage{algorithmic,algorithm}
\usepackage{enumerate}
\usepackage{cite}

\newtheorem{theorem}{Theorem}
\newtheorem{lemma}{Lemma}

\newtheorem{corollary}{Corollary}

\begin{document}
	
	\title{Wirelessly Powered Cell-free IoT: Analysis and Optimization}
	
	\author{ Xinhua Wang, Alexei Ashikhmin, \textit{Fellow, IEEE}, Xiaodong Wang, \textit{Fellow, IEEE}
		
		\thanks{Copyright (c) 2015 IEEE. Personal use of this material is permitted. However, permission to use this material for any other purposes must be obtained from the IEEE by sending a request to pubs-permissions@ieee.org.
		
Xinhua Wang is with the College of Electrical Engineering, Qingdao University, Qingdao, 266071 China (e-mail: xhwang@qdu.edu.cn).
         
Alexei Ashikhmin is with the Nokia Bell Labs, Murray Hill, NJ 07974 USA (e-mail: alexei.ashikhmin@nokia-bell-labs.com).
         
Xiaodong Wang is with the Electrical Engineering Department, Columbia University, New York, NY 10027 USA (e-mail: wangx@ee.columbia.edu).
			
		}
	}
	
	\maketitle
	
	\begin{abstract}
		In this paper, we propose a wirelessly powered Internet of Things (IoT) system based on the cell-free massive MIMO technology. In such a system, during the downlink phase, the sensors harvest radio-frequency (RF) energy emitted by the distributed access points (APs). During the uplink phase, sensors transmit data to the APs using the harvested energy. Collocated massive MIMO and small-cell IoT can be treated as special cases of cell-free IoT. We derive the tight closed-form lower bound on the amount of harvested energy, and the closed-form expression of SINR as the metrics of power transfer and data transmission, respectively. To improve the energy efficiency, we jointly optimize the uplink and downlink power control coefficients to minimize the total transmit energy consumption while meeting the target SINRs. Extended simulation results show that cell-free IoT outperforms collocated massive MIMO and small-cell IoT both in terms of the per user throughput for uplink, and the amount of energy harvested for downlink. Moreover, significant gains can be achieved by the proposed joint power control in terms of both per user throughput and energy consumption.
	\end{abstract}

	\section{Introduction}
	\IEEEPARstart{T}{he} Internet of Things (IoT) is envisioned as a promising technology which enables massively connected intelligent devices to share information and to coordinate decisions  \cite{A. Al-Fuqaha, L. D. Xu}. The concept of IoT has brought revolutionary applications in a wild range of domains including transportation, smart healthcare, environmental monitoring, smart home, and so on. However, the short battery life of the devices causes a bottleneck hampering the proliferation of IoT \cite{Z. Chu}.

     Wireless power transfer (WPT) has recently gained significant attention since it allows to prolong the lifetime of IoT and it is more controllable and reliable compared with ambient sources such as solar, wind, etc. \cite{S. H. Chae, D. S. Gurjar}. In wirelessly powered communication networks (WPCNs), the terminals first harvest RF energy from the WPT beacons, and then transmit information in the following time slots \cite{Y. Alsaba,T. D. Ponnimbaduge Perera}. This approach  can be extended to IoT networks with a large number of low power sensors.

    The main challenge of WPT is the low efficiency due to radio scattering and path loss \cite{X. Lu, J. Huang_a}. As effective counter measures, MIMO, and especially massive MIMO techniques, have been adopted in WPCNs \cite{T. A. Khan}, so that the sensors can harvest more energy since the RF energy becomes more concentrated. For massive MIMO based WPCN, Wu \textit{et al.} investigated the asymptotically optimal downlink power allocation strategy to maximize the uplink sum rate \cite{X. Wu}. The massive MIMO powered two-way and multi-way relay networks were investigated in \cite{X. Wang_a} and \cite{G. Amarasuriya}, respectively. However, the performance of cell-boundary terminals is still poor due to the heavy path-loss. The distributed antenna system (DAS) is adopted to reduce the path loss and improve the WPT efficiency. For distributed WPT system, Lee \textit{et al.} studied the effective channel training method for optimal energy beamforming with and without coordination \cite{S. Lee}. Kim \textit{et al.} proposed a joint time allocation and energy beamforming approach to maximize the energy efficiency of WPCN with DAS \cite{W. Kim}.

	Recently, cell-free massive MIMO wireless systems attracted  intensive research interests. In cell-free massive MIMO, a large number of access points (APs) are distributed over a large area. These APs collaboratively serve a large number of terminals using the same time-frequency resource \cite{H. Q. Ngo_a}, \cite{E. Nayebi}. In contrast to collocated (cellular) massive MIMO, cell-free massive MIMO is a user-centric architecture \cite{S. Buzzi}, since each terminal is served by the adjacent distributed APs. Compared with collocated massive MIMO, cell-free massive MIMO typically yields a high degree of macro-diversity and low path loss, since the service antennas are close to the sensors.
	Ngo \textit{et al.} derived the closed-form expressions of spectral efficiency and energy efficiency for the downlink cell-free massive MIMO system \cite{H. Q. Ngo_b}. To improve the spectral efficiency or energy efficiency, the precoding and power control are investigated in \cite{E. Nayebi} and \cite{L. D. Nguyen}. In a word, the cell-free massive MIMO can reap all benefits from DAS and massive MIMO. Recently first results on cell-free IoT (IoT based on cell-free massive MIMO) have been obtained in \cite{S. Rao}.
	
	\emph{\textbf{Motivation and Contribution}}:  It is intuitively clear that in cell-free IoT systems  the sensors can harvest more energy during the downlink power transfer phase and reduce the power consumption during the uplink data transmission phase. Motivated by such double-fold benefits, we consider a cell-free massive MIMO based IoT, in which some active sensors transmit signals to APs using the harvested energy during the downlink wireless power transfer.
	
	Our contributions in this work are two-fold:
	\begin{itemize}
		\item We propose the framework of wireless powered IoT based on cell-free massive MIMO. Collocated massive MIMO and small-cell IoT can be treated as special cases of cell-free IoT. We derive the tight closed-form lower-bound on the amount of harvested energy, and the closed-form expression of SINR for three systems (cell-free IoT, collocated massive MIMO, and small cell IoT), respectively. Numerical comparisions show that the cell-free IoT system has the best uplink and downlink performances. 
		\item  The uplink and downlink power control coefficients are jointly optimized to minimize the total energy consumption while meeting the predefined target SINR. The problem is equivalently decomposed into a linear optimization problem for uplink data transmission, and a quadratic optimization problem for downlink power transfer. Closed-form solutions to both problems are provided.
	\end{itemize}
	The remainder of this paper is organized as follows. In Section II we describe system model and outline our results. In Section III we derive expressions for uplink and downlink performances. In Section IV, we formulate and solve the joint power control problem. Simulation results are given in Section V. Finally in Section VI concludes the paper.
	
	\emph{\textbf{Notation}}: Throughout this paper, scalars and vectors are denoted by lowercase letters and boldface lowercase letters, respectively. Diag$({\textbf{a}})$ denotes a diagonal matrix with diagonal entries equal to the components of $\textbf{a}$. $\left|\cdot\right|$ and $\left\|\cdot\right\|$ represent the absolute value and the $\ell_2$ norm, respectively. $(\cdot)^H$ and $(\cdot)^{-1}$ denote the conjugate transpose and the inverse operation, respectively. $\left[{\bf{A}}\right]_{mm}$ returns the $m$-th diagonal element of ${\bf{A}}$. $\mathcal{CN}\left(\textbf{m},\textbf{R}\right)$ denotes the circularly symmetric complex Gaussian (CSCG) distribution with mean $\textbf{m}$ and covariance matrix $\textbf{R}$. $\mathbb{E}[\cdot]$ and var$\{\cdot\}$ stand for the expectation and variance operations, respectively.

	\section{System Model and Outline of Results}
	We consider a wirelessly powered IoT based on cell-free massive MIMO as shown in Fig. 1, in which \(L\) distributed APs serve a large number of sensors that are randomly located in a large area. Among them, there are \(K\) active sensors in a given period. We assume that APs know the active sensors which are indexed as $1, \ldots ,K$. Each AP is equipped with \(N\) antennas and each user has a single antenna. The channel coefficient between the $k$-th sensor and the $n$-th antenna of the \(l\)-th AP is denoted as  
	\[g_{(l,n),k}=\sqrt{\beta_{l,k}}h_{(l,n),k}, \] where \(\beta_{l,k}\) represents the large-scale fading and is assumed known, and $h_{(l,n),k}\sim \mathcal{CN}\left(0,1 \right) $ is the small-scale fading. Denote \(\pmb{g}_{(l,n)}\) as the channel vector between the \(n\)-th antenna of \(l\)-th AP and active sensors, and \({\pmb{g}}_{l,j}\) as the channel vector between the \(l\)-th AP and the \(j\)-th sensor, i.e.
	\[\pmb{g}_{(l,n)}=\left[g_{(l,n),1},\cdots,g_{(l,n),K}\right]^T ~\in \mathbb{C}^{K \times 1},\]
	and
	\[{\pmb{g}}_{l,j}=\left[{g}_{(l,1),j},\cdots,{g}_{(l,N),j}\right]^T~\in \mathbb{C}^{N \times 1}.\] 
	All APs connect to a Central Processing Unit (CPU) via a perfect back-haul network and collaboratively serve all users using the same time-frequency resource under TDD operation.
	\begin{figure}
\centering \scalebox{1}{\includegraphics[width=\columnwidth]{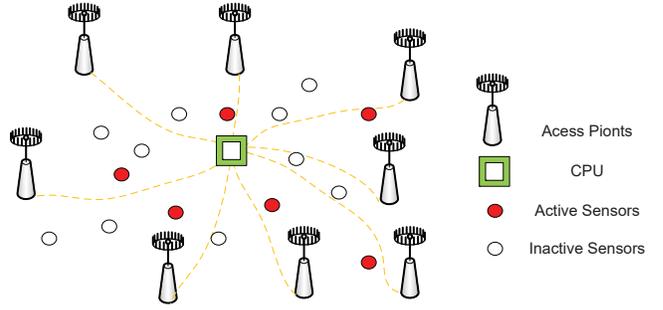}}
\centering \caption{Cell-free massive MIMO with distributed APs serving active sensors. }
\end{figure}

\begin{figure}
\centering \scalebox{1}{\includegraphics[width=\columnwidth]{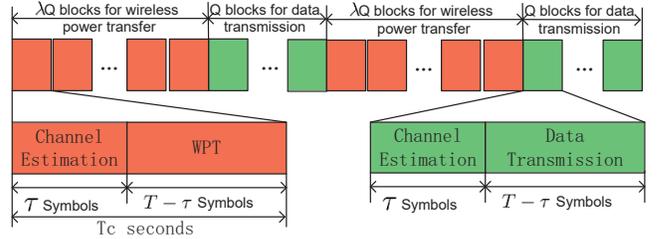}}
\centering \caption{The frame structure.}
\end{figure}
	As show in Fig. 2, we partition communication into periods, and each period includes \((\lambda+1)Q\) consecutive coherence time blocks. In each period, the \(K\) active users first harvest RF energy emitted by APs over \(\lambda Q\) time blocks, and next transmit data to APs in the remaining time blocks using the harvested energy. Each coherence time \(T_c\) block contains \(T\) OFDM symbols, in which \(\tau\) symbols are used for channel estimation, while the remaining symbols are used for WPT or data transmission.
    \subsection{System Model}
\subsubsection{Downlink WPT}

During the \(\tau\) symbols in each time slot, all \(K\) active sensors simultaneously transmit their pilot sequences to all APs for channel estimation. Let \(\pmb{\psi}_k \in \mathbb{C}^{\tau}\) with \(\left\|\pmb{\psi}_k\right\|^2=1\) be the pilot sequence of the \(k\)-th sensor. Denote \(\pmb{\Psi}=\left[\pmb{\psi}_{1},\cdots,\pmb{\psi}_{K}\right] \in \mathbb{C}^{\tau\times K}\), the received pilots \(\pmb{y}_{(l,n)} \in \mathbb{C}^{\tau}\) at the \(n\)-th antenna of the \(l\)-th AP is given by
	\begin{align}	\label{equ:sys1}　\pmb{y}_{(l,n)}&=\sqrt{\tau\rho_p}\sum_{k=1}^{K}{g}_{(l,n),k}\pmb{\varphi}_k+\pmb{w}_{(l,n)}\nonumber \\
	&=\sqrt{\tau\rho_p}\pmb{\Psi}\pmb{g}_{(l,n)}+\pmb{w}_{(l,n)},
	\end{align}
	where \(\pmb{w}_{(l,n)}\sim \mathcal{CN}(0,\pmb{I})\) is the additive noise,  and \(\rho_p\) is the normalized pilot transmit power. Given \(\pmb{y}_{(l,n)} \), the channel estimate \(\hat{\pmb{g}}_{(l,n)}\) is obtained by using the linear minimum mean square error (LMMSE) method.
	
	During the remaining symbols in each time slot, the APs use the estimated channels to conduct conjugate beamforming, and simultaneously transmit signals to all sensors. Denote by \(\eta_{l,j} \) the power control coefficients of the \(l\)-th AP for the \(j\)-th sensor, and by \(q_j \sim \mathcal{CN}(0,1)\) the symbol intended for this sensor. The received signal at the \(k\)-th sensor is
	\begin{align}\label{equ:sys2}
	z_k&=\sum_{l=1}^L\pmb{g}^T_{l,k}\pmb{x}_{l}+v_k,
	\end{align}
	where \(v_k\sim \mathcal{CN}(0,~1)\) is the additive noise at the \(k\)-th sensor, and \( \pmb{x}_{l}=\sqrt{\rho_d}\sum_{j=1}^K \sqrt{\eta_{l,j}}{\hat{\pmb{g}}^{*}_{l,j}} q_j\) is the transmitted signal from the \(l\)-th AP with
	\begin{align}\label{equ:sys3}
	 \Xi_l=\mathbb{E}\left[\left\|\pmb{x}_l\right\|^2\right]\leq N\rho_d,  
	\end{align}
	 where \(N\rho_d\) is the maximum transmit power of each AP. Thus, the total energy consumption during the \(\lambda Q\) downlink WPT time blocks is
   	\begin{align}\label{equ:sys4}
 \Xi_{\rm tr}=(1-\frac{\tau}{T}) \lambda Q \sum_{l=1}^L \Xi_l,	\end{align}
	while the harvested energy of the \(k\)-th sensor during the \(\lambda Q\) WPT time blocks can be expressed as
		 \begin{align}\label{equ:sys5}
\mathcal{E}_{k}&=(1-\frac{\tau}{T}) \lambda Q\zeta\mathbb{E}\left[\left|z_k\right|^2\right],
  \end{align}
where \(\zeta \in [0,~1]\) is the energy conversion efficiency.
	\subsubsection{Uplink Data Transmission}
	During the \(\tau\) symbols in each time slot, channel estimation is performed in the same way as the downlink WPT case. During the remaining symbols in each time slot, \(K\) users simultaneously transmit their data to all APs. Let \(\rho_u\) be the maximum normalized transmit power of each sensor. Let \(\xi_j\in [0,~1]\) be the power control coefficient, and \(s_j\) be the data symbol of the \(j\)-th user with \(\mathbb{E}\left[|s_j|^2\right]=1\). Then, the received signal \(\pmb{r}_{l} \in \mathbb{C}^N\) at the \(l\)-th AP is
    \begin{align}\label{equ:sys6}
    \pmb{r}_{l}=\sqrt{\rho_u}\sum_{j=1}^{K}\sqrt{\xi_j}\pmb{g}_{l,j}s_j+\pmb{n}_{l},
    \end{align}
    where \(\pmb{n}_l\sim \mathcal{CN}({\pmb 0},~{\pmb I}_N)\) is the additive noise. To detect symbol \(s_k\),  the \(l\)-th AP computes \(\hat{\pmb{g}}^H_{lk}\pmb{y}_{l}\) and sends it to the CPU. The CPU employs the equal gain combining (EGC) to detect \(s_k\) as follows
	\begin{align}\label{equ:sys7}
	\hat{s}_k&=\sum\nolimits_{l=1}^L\hat{\pmb{g}}_{l,k}^{H}\pmb{r}_{l}\nonumber\\
	&=\underbrace{\sqrt{\rho_u\xi_k}\sum\nolimits_{l=1}^L\mathbb{E}\left[\hat{\pmb{g}}_{l,k}^{H}\pmb{g}_{l,k}\right]}_{\mathcal{A}_1}s_k\nonumber \\&+\underbrace{\sqrt{\rho_u\xi_k}\sum\nolimits_{l=1}^L\left(\hat{\pmb{g}}_{l,k}^{H}\pmb{g}_{l,k}-\mathbb{E}\left[\hat{\pmb{g}}_{l,k}^{H}\pmb{g}_{l,k}\right]\right)}_{\mathcal{A}_2}s_k\nonumber \\&+\underbrace{\sum_{j\neq k}^{K}\sqrt{\rho_u\xi_j}\sum\nolimits_{l=1}^L\hat{\pmb{g}}_{l,k}^{H}\pmb{g}_{l,j}}_{\mathcal{A}_3}s_j+\underbrace{\sum\nolimits_{l=1}^L\hat{\pmb{g}}_{l,k}^{H}\pmb{n}_{l}}_{\mathcal{A}_4},
	\end{align}
 where \(\mathcal{A}_1\) is the desired signal, \(\mathcal{A}_2\), \(\mathcal{A}_3\), and \(\mathcal{A}_4\) are the beamforming uncertainty, inter-user interference due to the non-orthogonality of the pilots, and noise, respectively. It is not difficult to show that \(\mathcal{A}_1\), \(\mathcal{A}_2\), \(\mathcal{A}_3\), and \(\mathcal{A}_4\) are uncorrelated. Hence according to \cite{ B. Hassibi}, the worst case is the AWGN channel with the effective noise \(\mathcal{A}_2+\mathcal{A}_3+\mathcal{A}_4\). Thus, similarly as in \cite{H. Q. Ngo_a}, the capacity of the \(k\)-th sensor is lower bounded by 
 \begin{align}
    C_k=\log_2(1+\Gamma_k) ~~\mbox{bits/s/Hz},
\end{align}
with the effective SINR
 \begin{align} \label{equ:sys8}	
	\Gamma_k=\frac{|\mathcal{A}_1|^2 }{\mathbb{E}\left[|\mathcal{A}_2|^2\right]+\mathbb{E}\left[|\mathcal{A}_3|^2\right]+\mathbb{E}\left[|\mathcal{A}_4|^2\right]},
	\end{align}
where the expectation is with respect to the small scale fading. In addition, the energy consumption of the \(k\)-th sensor during successive \(Q\) time blocks for data transmission is
 \begin{align}\label{equ:sys9}
E_{k}(\xi_k)=(1-\frac{\tau}{T})Q\rho_u \xi_k.
\end{align}

\subsection{Outline of Results}
To evaluate the performance of the cell-free IoT, a collocated massive MIMO system and a small-cell system are also considered as benchmarks for comparison. The collocated massive MIMO can be treated as a special case of cell-free IoT, where all \(L\) APs are collocated, which implies \(\beta_{l,k}=\beta_{k}, \forall l.\) For the small-cell system, we assume that user \(k\) is served by only one AP that has the largest \(\beta_{l,k}\) coefficient. We define the following binary association coefficient
\[\delta_{l,j}=\left\lbrace\begin{aligned}&1,&j\mbox{-th~ sensor~is~associated~with~the}~l\mbox{-th~AP}, \nonumber\\
&0,& {\rm otherwise.}\qquad\qquad\qquad\qquad\qquad\qquad\quad \end{aligned}\right.\]
Then, the received signal at the \(k\)-th sensor during downlink WPT phase (corresponding to (\ref{equ:sys2}) of cell-free IoT) is
			\begin{align*}
		   	z_k^{\rm sc}&=\sum_{l=1}^L\delta_{l,k}\pmb{g}^T_{l,k}\pmb{x}^{\rm sc}_{l}+v_k,\nonumber
			\end{align*}
where \(\pmb{x}_{l}^{sc}=\sqrt{\rho_d}\sum_{j=1}^K \sqrt{\delta_{l,j}\eta_{l,j}}{\hat{\pmb{g}}^{*}_{l,j}} q_j \) is the transmitted signal at the \(l\)-th AP. Similarly as cell-free IoT, during uplink data transmission, the estimate of \(s_k\) is 
\begin{align*}
    	\hat{s}^{\rm sc}_k&=\sum_{l=1}^L\delta_{l,k}\hat{\pmb{g}}_{l,k}^{H}\pmb{y}_{l}.
    	\end{align*}
Hence, the small-cell system can also be treated as a special case of cell-free IoT with \(\hat{\pmb{g}}_{l,k}=\delta_{l,k}\hat{\pmb{g}}_{l,k}\).

In Section III, we derive the tight closed-form lower-bound of \({\mathcal E}_{k}\) in (\ref{equ:sys5}) and the closed-form expression of \(\Gamma_k\) in (\ref{equ:sys8}) as the metrics of WPT and data transmission respectively for the three systems. Numerical results reveal that cell-free massive MIMO achieves higher \({\mathcal E}_{k}\) and \(\Gamma_k\) given the same power control coefficients. This is because, compared with collocated massive MIMO, the cell-free massive MIMO can achieve more macro-diversity since the sensors are closer to APs; and compared with small-cell, the cooperation between different APs leads to higher array gain.  

Then in Section IV, we jointly optimize the downlink and uplink power control coefficients \({\pmb \eta}, {\pmb \xi} \), and the WPT duration \(\lambda\) to further improve the efficiency of the cell-free IoT. We aim to minimize the energy consumption of APs \(\Xi_{\rm tr}\) in (\ref{equ:sys4}) while meeting a given target SINR during data transmission supported by the harvested energy. 
  	
\section{Performance Analysis}
	In this section, we derive tight closed-form lower-bounds on \(\mathcal{E}_k\) in (\ref{equ:sys5}), and the closed-form expressions of \(\Gamma_k\) in (\ref{equ:sys8}) for cell-free massive MIMO, collocated massive MIMO, and small-cell systems.
	\subsection{LMMSE Channel Estimation}
According to (\ref{equ:sys1}), we have
	\[
	\begin{split}
	&\mathbb{E}\left[ \pmb{y}_{(l,n)}\pmb{y}_{(l,n)}^H\right]=\mathbb{E}\left[\left( \sqrt{\tau\rho_p}\pmb{\Psi}\pmb{g}_{(l,n)}+\pmb{w}_{(l,n)}\right)\times\right.\cr
	&\left.\left( \sqrt{\tau\rho_p}\pmb{g}_{(l,n)}^H\pmb{\Psi}^H+\pmb{w}_{(l,n)}^H\right) \right] =\tau \rho_p \pmb{\Psi}\pmb{D}_l\pmb{\Psi}^H+\pmb{I}, \qquad
	\end{split}\]and
	\[\begin{split}
	&\mathbb{E}\left[ \pmb{g}_{(l,n)}\pmb{y}_{(l,n)}^H\right] =\mathbb{E}\left[\pmb{g}_{(l,n)} \left( \sqrt{\tau\rho_p}\pmb{g}_{(l,n)}^H\pmb{\Psi}^H+\pmb{w}_{(l,n)}^H\right) \right] \cr
	&=\mathbb{E}\left[\sqrt{\tau\rho_p}\pmb{g}_{(l,n)}\pmb{g}_{(l,n)}^H\pmb{\Psi}^H+\pmb{g}_{(l,n)}\pmb{w}_{(l,n)}^H \right]=\sqrt{\tau\rho_p}\pmb{D}_l\pmb{\Psi}^H, 
	\end{split}\]
	where \(\pmb{D}_l\overset{}{=}\mathbb{E}\left[\pmb{g}_{(l,n)}\pmb{g}_{(l,n)}^H\right]=\text{diag}\left(\beta_{l,1},\cdots,\beta_{l,k}\right).\) Thus, the LMMSE channel estimate of \({\pmb{g}}_{(l,n)}\) is
	\begin{align}\label{Per_A.1}　
	\hat{\pmb{g}}_{(l,n)}&=\mathbb{E}\left[ \pmb{g}_{(l,n)}\pmb{y}_{(l,n)}^H\right]\left( \mathbb{E}\left[ \pmb{y}_{(l,n)}\pmb{y}_{(l,n)}^H\right] \right)^{-1} \pmb{y}_{(l,n)}, \nonumber \\
    &=\sqrt{\tau\rho_p}\pmb{D}_l\pmb{\Psi}^H\left(\tau\rho_p\pmb{\Psi}\pmb{D}_l\pmb{\Psi}^H+\pmb{I}\right) ^{-1}\pmb{y}_{(l,n)}\nonumber\\
    &={\pmb A}_l^H{\pmb y}_{(l,n)},
	\end{align}
	where
$$
\pmb{A}_{l}=\sqrt{\tau\rho_p}\left(\tau\rho_p\pmb{\Psi}\pmb{D}_l\pmb{\Psi}^H+\pmb{I}\right) ^{-1}\pmb{\Psi}\pmb{D}_l.
$$
Thus, we have
\begin{align}\label{Per_A.2}
	\mathbb{E}\left[ \hat{\pmb{g}}_{(l,n)}\hat{\pmb{g}}_{(l,n)}^H\right] =\sqrt{\tau\rho_p}\pmb{D}_l\pmb{\Psi}^H\pmb{A}_{l} .
	\end{align}

The estimated channel \(\hat{\pmb{g}}_{(l,n)}\) includes \(K\) Gaussian distributed variables with
	\begin{align}\label{Per_A.3}
	 \gamma_{l,k}&=\mathbb{E}\left[\left|{\hat g}_{(l,n),k}\right|^2\right]=\left[\mathbb{E}\left(\hat{\pmb{g}}_{(l,n)}\hat{\pmb{g}}_{(l,n)}^H\right)\right]_{kk}\nonumber \\
  &=\sqrt{\tau\rho_p}\beta_{l,k}\pmb{\psi}_k^H\pmb{a}_{l,k}=\tau\rho_p\beta^2_{l,k}\pmb{\psi}_k^H\pmb{Z}_l^{-1}\pmb{\psi}_k,
	\end{align}
\begin{align}
&\mbox{ where }~&\pmb{Z}_l&=\tau\rho_p\pmb{\Psi}\pmb{D}_l\pmb{\Psi}^H+\pmb{I}, \qquad \qquad\qquad\qquad\label{Per_A.4}\\
&\mbox{ and ~~~}&\pmb{a}_{l,k}&=\sqrt{\tau\rho_p}\beta_{l,k}\pmb{Z}_l^{-1}\pmb{\psi}_k\qquad\qquad \qquad\qquad \label{Per_A.5}
\end{align}
is the \(k\)-th column of \(\pmb{A}_{l}\). It is also useful to write explicitly that
\begin{align}\label{Per_A.6}
{\hat g}_{(l,n),k}=\pmb{a}^H_{l,k}\left(\sqrt{\tau\rho_p}\sum_{i=1}^{K}{g}_{(l,n),i}\pmb{\varphi}_i+\pmb{w}_{(l,n)}\right).
\end{align}
We have the following lemma for the estimator in (\ref{Per_A.1}).
 \begin{lemma}
The estimates of channel vectors between different APs and the same sensor are uncorrelated, i.e.,
 $$
 {\rm cov}\left[\hat{\pmb g}_{l,k},~\hat{\pmb g}_{m,k}\right]={\pmb 0},~m,l \in \left\{1,\cdots,L\right\},m \neq l,~k=1,\ldots,K.
  $$
  Moreover, the corresponding norms are also uncorrelated, i.e.,
  $$
  {\rm cov}\left[\left\|\hat{\pmb g}_{l,k}\right\|^2,~\left\|\hat{\pmb g}_{m,k}\right\|^2\right]=0.
  $$
 \end{lemma}
\begin{IEEEproof}
See Appendix A.
 \end{IEEEproof}
 
	\subsection{Results for Cell-free IoT}
	\subsubsection{Downlink Power Transfer}
	Let \(\tilde{\pmb{g}}_{lk}=\pmb{g}_{lk}-\hat{\pmb{g}}_{lk}\) be the channel estimation error. The received signal at the \(k\)-th user in (\ref{equ:sys2}) can be rewritten as
	
	\begin{align}
	&z_k =\mathcal{S}_{k1}+\mathcal{S}_{k2}+\mathcal{S}_{k3},\label{Per_B.1}\\
{\rm where}~~~~&\mathcal{S}_{k1}=\sqrt{\rho_d}\sum\nolimits_{l=1}^L\sqrt{\eta_{l,k}}\hat{\pmb{g}}^T_{l,k}{\hat{\pmb{g}}^{*}_{l,k}} q_k,\nonumber \\
	&\mathcal{S}_{k2}=\sqrt{\rho_d}\sum\nolimits_{l=1}^L\sqrt{\eta_{l,k}}\tilde{\pmb{g}}^T_{l,k}{\hat{\pmb{g}}^{*}_{l,k}} q_k,\nonumber\\
	{\rm and~~~} &\mathcal{S}_{k3}=\sqrt{\rho_d}\sum\nolimits_{l=1}^L\sum\nolimits_{j\neq k}^K \sqrt{\eta_{l,j}}\pmb{g}^T_{l,k}{\hat{\pmb{g}}^{*}_{l,j}} q_{j }+v_k.\nonumber
	\end{align}

The amount of energy harvested by the \(k\)-th user during \(\lambda Q\)  successive time blocks can be expressed as
 \begin{align*}
&\mathcal{E}_{k}=(1-\frac{\tau}{T}) \lambda Q\zeta\mathbb{E}\left[\left|\mathcal{S}_{k1}+\mathcal{S}_{k2}+\mathcal{S}_{k3}\right|^2\right]\\
&=(1-\frac{\tau}{T}) \lambda Q\zeta\mathbb{E}
\left[\left|\mathcal{S}_{k1}
\right|^2+\left|\mathcal{S}_{k2}+\mathcal{S}_{k3}\right|^2  +2\Re\{\mathcal{S}_{k1}\left(\mathcal{S}_{k2}+\mathcal{S}_{k3}\right)\}\right].
  \end{align*}
Note that \(\mathcal{S}_{k1}\), \(\mathcal{S}_{k2}\), and \(\mathcal{S}_{k3}\) are uncorrelated since we assume that downlink symbols for different users are uncorrelated. Thus, we have
   $\mathbb{E}\left[2\Re\{\mathcal{S}_{k1}\left(\mathcal{S}_{k2}+\mathcal{S}_{k3}\right)\}\right]=0$, and this allows us to get the following lower bound for \(\mathcal{E}_{k}\) as (\ref{Per_B.2}) shown at the top of next page,
    where step (a) is obtained according to Lemma 1 and  \(\tilde{\eta}_{l,k}={\eta}_{l,k}{\gamma}_{l,k}\) should satisfy
\begin{align*}
\sum_{k=1}^K \tilde{\eta}_{lk} \leq 1,~\mbox{ for any AP } l=1,\ldots,L.
\end{align*}
 according to (\ref{equ:sys3}) and (\ref{Per_A.3}), where the constant \(\gamma_{l,k}\) is given in (\ref{Per_A.3}), which essentially is the estimate of \(\beta_{l,k}\).
	\begin{figure*}
	    \begin{flalign}\label{Per_B.2}
 &\mathcal{E}_{k}\geq \tilde{\mathcal{E}}_{k}=(1-\frac{\tau}{T}) \lambda Q\zeta\mathbb{E}\left[\left|\mathcal{S}_{k1}\right|^2\right]= (1-\frac{\tau}{T}) \lambda Q\zeta \rho_d \sum\nolimits_{l=1}^{L}\sum\nolimits_{m=1}^{L}\mathbb{E}\left[\sqrt{\eta_{l,k}\eta_{m,k}}\hat{\pmb{g}}^T_{l,k}{\hat{\pmb{g}}^{*}_{l,k}}\hat{\pmb{g}}^T_{m,k}{\hat{\pmb{g}}^{*}_{m,k}}\right]\nonumber\\
&=(1-\frac{\tau}{T}) \lambda Q\zeta \rho_d \sum\nolimits_{l=1}^{L}\mathbb{E}\left[\eta_{l,k}\left\|\hat{\pmb{g}}_{l,k}\right\|^4\right]+(1-\frac{\tau}{T}) \lambda Q\zeta \rho_d \sum\nolimits_{l=1}^{L}\sum\nolimits_{m\neq l}^{L}\mathbb{E}\left[\sqrt{\eta_{l,k}\eta_{m,k}}\left\|\hat{\pmb{g}}_{l,k}\right\|^2\left\|\hat{\pmb{g}}_{m,k}\right\|^2\right]\nonumber\\
&\overset{(a)}{=} (1-\frac{\tau}{T}) \lambda Q\zeta N(N+1)\rho_d \sum\nolimits_{l=1}^{L}\left(\eta_{l,k}\gamma^{2}_{l,k}\right)+(1-\frac{\tau}{T}) \lambda Q\zeta N^2\rho_d \sum\nolimits_{l=1}^{L}\sum\nolimits_{m\neq l}^{L}\left (\sqrt{\eta_{l,k}\eta_{m,k}}\gamma_{l,k}\gamma_{m,k}\right)\nonumber\\
&= (1-\frac{\tau}{T}) \lambda Q\zeta N^2\rho_d \left(\sum\nolimits_{l=1}^{L}\sqrt{\eta_{l,k}}\gamma_{l,k}\right)^2+(1-\frac{\tau}{T}) \lambda Q\zeta N\rho_d \sum\nolimits_{l=1}^{L}\left(\sqrt{\eta_{l,k}}\gamma_{l,k}\right)^2\nonumber\\
&=(1-\frac{\tau}{T}) \lambda Q\zeta N\rho_d\left[N\left(\sum\nolimits_{l=1}^{L}\sqrt{\tilde{\eta}_{l,k}\gamma_{l,k}}\right)^2+\sum\nolimits_{l=1}^{L}{\tilde{\eta}_{l,k}\gamma_{l,k}}\right].
	\end{flalign}
    \hrule
	\end{figure*}

	\subsubsection{Uplink Data Transmission}
Using the method in \cite{S. Rao}, we get the following closed-form expression of the SINR given in (\ref{equ:sys8}) which is a function of the large-scale fading coefficients and the pilot sequences.
	\begin{theorem}
	The effective SINR of the \(k\)-th sensor in cell-free massive MIMO with LMMSE channel estimation and EGC receiver is
    \begin{align}\label{Per_B.3}	
    \Gamma_k=\frac{\mathcal{D}_{k}\xi_k}{\mathcal{U}_{k}\xi_k+\sum_{j\neq k}\mathcal{I}_{kj}\xi_j+\mathcal{N}_{k}
	},
    \end{align}
   where
   \begin{align}
    \mathcal{D}_{k}&=\rho_uN\left(\sum\nolimits_{l=1}^L\gamma_{l,k} \right)^2,
    \mathcal{U}_{k}=\sum\nolimits_{l=1}^L \rho_u\gamma_{l,k}\beta_{l,k},\nonumber \\
    \mathcal{N}_{k}&=\sum\nolimits_{l=1}^L\gamma_{l,k}, \mbox{and} \nonumber\\
   \mathcal{I}_{kj}&=\rho_u\sum\nolimits_{l=1}^L\beta_{l,j}\left\|\pmb{a}_{l,k}\right\|^2+\tau\rho_u\rho_pN\left(\sum\nolimits_{l=1}^L\beta_{l,j}\pmb{\psi}_j^H\pmb{a}_{l,k}
   \right)^2\nonumber \\
    &+\tau\rho_u\rho_p\sum\nolimits_{l=1}^L\sum\nolimits_{i=1}^{K}\beta_{l,j}\beta_{l,i}\left(\pmb{\psi}_i^H\pmb{a}_{l,k}\right)^2\nonumber
   \end{align}
	\end{theorem}
	
\begin{IEEEproof}
See Appendix B.
\end{IEEEproof}

\subsection{Results for Collocated Massive MIMO and Small-cell IoT}
The collocated massive MIMO is a special case with \(\beta_{l,k}=\beta_{k}\), \(\gamma_{l,k}=\gamma_{k}\), and \(\tilde{\eta}_{l,k}=\tilde{\eta}_{k}\). So, we have the following corollary.
\begin{corollary}
 For collocated massive MIMO, the amount of energy harvested by the \(k\)-th user in \(\lambda Q\) successive time blocks is lower bounded as 
 \begin{align*}
{\mathcal{E}}^{\rm cm}_{k}\geq \tilde{\mathcal{E}}^{\rm cm}_{k}=(1-\frac{\tau}{T}) \lambda Q\zeta\rho_dLN\left(LN+1\right){\tilde{\eta}_{k}{\gamma}_{k}}.
	\end{align*}
The effective SINR of the \(k\)-th sensor during data transmission phase is given by
 \begin{align*}
    	\Gamma^{\rm cm}_{k}=
\frac{\mathcal{D}_{k}^{\rm cm}\xi_k}{\mathcal{U}_{k}^{\rm cm}\xi_k+\sum_{j\neq k}\mathcal{I}_{kj}^{\rm cm}\xi_j+\mathcal{N}^{\rm cm}_{k}
	}
    \end{align*}
    where \(\mathcal{D}_{k}^{\rm cm}=\rho_uLN\left(\gamma_{k} \right)^2\),  \(	\mathcal{U}_{k}^{\rm cm}= \rho_u\gamma_{k}\beta_{k}\), \(	\mathcal{N}_{k}^{\rm cm}=\gamma_{k} \),   \(\mathcal{I}_{kj}^{\rm cm}=\rho_u\beta_{j}\left\|\pmb{a}_{k}\right\|^2+\tau\rho_u\rho_p\sum_{i=1}^{K}\beta_{j}\beta_{i}\left(\pmb{\psi}_i^H\pmb{a}_{k}\right)^2+\tau\rho_u\rho_pLN\left(\beta_{j}\pmb{\psi}_j^H\pmb{a}_{k}\right)^2\).
\end{corollary}

Moreover, the small-cell IoT is also a special case with  \(\hat{\pmb{g}}_{l,k}=\delta_{l,k}\hat{\pmb{g}}_{l,k}\) and \(\gamma_{l,k}=\delta_{l,k}\gamma_{l,k}\). Substituting them into (\ref{Per_B.2}) and (\ref{Per_B.3}), we have the following corollary.
\begin{corollary}
 For small-cell IoT, the amount of energy harvested by the \(k\)-th user in \(\lambda Q\) successive time blocks is lower bounded as
			\begin{align}\label{perf11}
			{\mathcal{E}}^{sc}_{k}\geq \tilde{\mathcal{E}}^{sc}_{k}
			&=(1-\frac{\tau}{T}) \lambda Q\zeta\rho_dN(N+1)\sum_{l=1}^L{\delta_{lk}\tilde{\eta}_{lk}{\gamma}_{lk}}.
			\end{align}
The effective SINR of the \(k\)-th sensor during data transmission phase is given by
\begin{align*}
    \Gamma^{sc}_{k}=\frac{\mathcal{D}_{k}^{sc}\xi_k}{\mathcal{U}_{k}^{sc}\xi_k+\sum_{j\neq k}\mathcal{I}_{kj}^{sc}\xi_j+\mathcal{N}^{sc}_{k}
	}
    \end{align*}
   where
   \begin{align*}
   &\mathcal{D}_{k}^{sc}=\rho_uN\sum\limits_{l=1}^L\delta_{l,k} \gamma^2_{l,k},
   \mathcal{U}_{k}^{sc}=\sum\limits_{l=1}^L \rho_u\delta_{l,k}\gamma_{l,k}\beta_{l,k},\nonumber \\
   &\mathcal{N}_{k}^{sc}=\sum\limits_{l=1}^L \delta_{l,k}\gamma_{l,k}, \mbox{ and } \nonumber \\
   &\mathcal{I}_{kj}^{sc}=\rho_u\left[\tau\rho_p\sum\limits_{l=1}^L\sum\limits_{i=1}^{K}\delta_{l,k}\beta_{l,j}\beta_{l,i}\left(\pmb{\psi}_i^H\pmb{a}_{l,k}\right)^2
  \right.\nonumber \\&\left.+ \sum\limits_{l=1}^L\delta_{l,k}\beta_{l,j}\left\|\pmb{a}_{l,k}\right\|^2+\tau\rho_pN\left(\sum\limits_{l=1}^L\delta_{l,k}\beta_{l,j}\pmb{\psi}_j^H\pmb{a}_{l,k}\right)^2\right].
   \end{align*}
\end{corollary}
	\section{Joint Downlink-Uplink Power Control}
	 To improve the energy efficiency of cell-free IoT, we aim to minimize the total energy consumption of APs \(\Xi_{\rm tr}\) in (\ref{equ:sys4}) while meeting a given target SINR by jointly optimize the uplink power control coefficients \(\pmb \xi\) and the downlink energy allocation \({\pmb \mu}=\lambda\tilde{\pmb \eta}\) with \(\mu_{l,k}=\lambda \tilde{\eta}_{l,k}\). Then, we determine the normalized downlink power control coefficients \(\tilde{\pmb \eta}\) through minimizing the WPT duration \(\lambda\). To prolong the lifetime of IoT, the amount of harvested energy in each period of each sensor should satisfy
	 
\begin{align}\label{joint_1}
    \tilde{\mathcal E}_{k}\geq {E}_{k}(\xi_k)+E_0,~ \forall k,
\end{align}
 where \({E}_{k}(\xi_k)\) is given by (\ref{equ:sys9}), and \(E_0\) is a constant which can satisfy the basic energy consumption. \(\tilde{\mathcal E}_{k}\) given in (\ref{Per_B.2}) can be rewritten as a function of \(\pmb{\mu}\) as follows

\begin{small} \begin{align}\label{equvalents1}
    \tilde{\mathcal{E}}_k({\pmb \mu})=(1-\frac{\tau}{T})  Q\zeta N\rho_d\left[N\left(\sum\limits_{l=1}^{L}\sqrt{{\mu}_{l,k}\gamma_{l,k}}\right)^2+\sum\limits_{l=1}^{L}{{\mu}_{l,k}\gamma_{l,k}}\right].
\end{align}
\end{small}
According to (\ref{equ:sys3}) and (\ref{equ:sys4}) and the definition of \({\pmb x}_l\), the total energy consumption of APs can be rewritten as
\begin{align}\label{joint_2}
   \Xi_{\rm tr}=Q (1-\frac{\tau}{T}) \rho_dN\sum_{l=1}^L\sum_{k=1}^K{\mu}_{l,k}. 
\end{align}
The joint optimization problem is then
\begin{align}
	{\bf P0}:  \mathop {\min }\limits_{\pmb{\xi},\pmb{\mu}} &\quad  \Xi_{\rm tr} \nonumber\\
	{s.t}. & \quad  \tilde{\mathcal{E}}_{k}({\pmb \mu})\geq {E}_{k}({\xi}_k)+C_0,~ \forall k,\nonumber\\
    &\quad {\Gamma}_{k}\left(\pmb{\xi}\right)\geq \Delta_k, ~ \forall k,\label{P1.1} \\
	&\quad 0 \leq {{\xi}_k}\leq 1, ~\forall k,\label{P1.2}\\
    	& \quad  {\pmb{\mu}}\geq 0,
	\end{align}
	where \(\Delta_k\) is a given target SINR value during the data transmission. Next we show that {\bf P0} can be equivalently decomposed into the following two problems.
\begin{align}
{\bf P1}: &	\mathop{ \min}_{\pmb{\xi}} \quad\sum_{k=1}^{K}  {E}_{k}({\xi}_k)\nonumber \\
	{s.t}. &\quad {\Gamma}_{k}\left(\pmb{\xi}\right)=\frac{\mathcal{D}_{k}\xi_k}{\mathcal{U}_{k2}\xi_k+\sum_{j\neq k}\mathcal{I}_{kj}\xi_j+\mathcal{N}_{k}
	}\geq \Delta_k, \forall k ,\nonumber\\
	&\quad 0 \leq {{\xi}_k}\leq 1,  \forall k,  \nonumber
		\end{align}
		and
\begin{align}    
    {\bf P2}: \mathop {\min }\limits_{\pmb{\mu}} &\quad \Xi_{\rm tr}\qquad \qquad \nonumber  \\
	{s.t}. & \quad \tilde{\mathcal{E}}_{k}({\pmb \mu})\geq {E}_{k}({\xi}_k)+E_0 ,~\forall k, \nonumber \\
    & \quad  {\pmb{\mu}}\geq 0.
	\end{align}
{\bf P1} is minimization of the total energy consumption \(\sum_{k=1}^K {E}_{k}(\xi_k)\) subject to the target SINR constraint $\Delta_k$ for the uplink data transmission, and {\bf P2} is minimization of the total energy consumption given the target harvested energy constraints for the downlink WPT.
\begin{theorem}
Solving {\bf P0} is equivalent to solving {\bf P1} and {\bf P2} in sequence.
\end{theorem}
\begin{IEEEproof}
 From Theorem 3 below, the optimal solution \(\pmb{\xi}^*\) to {\bf P1} is the point that can simultaneously minimize \({E}_{k}(\xi_k)\) for all \(k\) under the constraints of (\ref{P1.1}) and (\ref{P1.2}). That is, for any point \({\pmb \xi}\in \mathbb{P}\) with \(\mathbb{P}\) being the feasibility region defined by (\ref{P1.1}) and (\ref{P1.2}), we have 
\begin{align}
 {E}_{k}({\xi}_k) \geq {E}_{k}({\xi}_k^{*}),  k=1,\cdots, K.
\end{align}
Denote the optimal solution to {\bf P0} as \(({\pmb \xi}^{\#}, {\pmb \mu}^{\#})\). It is noted that \(\Xi_{\rm tr}\) and \(\tilde{\mathcal{E}}_k({\pmb \mu})\) are monotonically increasing functions w.r.t \({\mu}_{l,k}, \forall l, k\). Thus, for \({\pmb \xi}^{\#} \neq {\pmb \xi}^{*}\), we can further reduce \({\mu}_{l,k}, \forall l\) when \(
 {E}_{k}({\xi}^{\#}_k) > {E}_{k}({\xi}_k^{*}),\)
and get a new solution \(({\pmb \xi}^{*} ,{\pmb \mu}^{*})\) with \({\pmb \mu}^{*}\preceq {\pmb \mu}^{\#}\) which can further minimize the objective function \(\Xi_{\rm tr}\). Hence, the optimal solution to {\bf P0} can be achieved only when \({\pmb \xi}= {\pmb \xi}^{*}\), which implies that solving {\bf P0} is equivalent to solving {\bf P1} and {\bf P2} in sequence.
\end{IEEEproof}
In what follows, we discuss methods for solving {\bf P1} and {\bf P2}, respectively. 
	\subsection{Closed-form Optimal Solution to {\bf P1}}
Define the following \(K\times K\) matrix
\[
		\pmb W
		=\begin{bmatrix}
		\mathcal{D}_{1}-\Delta_1\mathcal{U}_{1} & -\Delta_1\mathcal{I}_{12}& \dots  & -\Delta_1\mathcal{I}_{1K} \\
		- \Delta_2\mathcal{I}_{21}&\mathcal{D}_{2}-\Delta_2\mathcal{U}_{2}& \dots  & -\Delta_2\mathcal{I}_{2K}\\
		\vdots & \vdots & \ddots & \vdots \\
		-\Delta_K\mathcal{I}_{K1} & -\Delta_K\mathcal{I}_{K2} & \dots  & \mathcal{D}_{K}-\Delta_K\mathcal{U}_{K}
		\end{bmatrix}.
		\]
We have the following result.
\begin{theorem}
If {\bf P1} is feasible with \(\mathbb{P}\neq \varnothing\), and \(\pmb W\) is invertiable, then the optimal solution \({\pmb \xi}^*=(\xi_1^*, \ldots, \xi_K^*)\) of {\bf P1} is given by
\begin{align}\label{Uplink_2}
      \pmb{\xi}^{*}=\pmb{W}^{-1}\pmb{b},
      \end{align}
where \(\pmb {b}=\left[\Delta_1\mathcal{N}_{1},\Delta_2\mathcal{N}_{2},\cdots,\Delta_K\mathcal{N}_{K}\right]^T
		\). In addition, \(\pmb{\xi}^{*}\) simultaneously minimizes the energy consumption for each sensor subject to the target SINR constraints, i.e., 
\begin{align}\label{Th1}
 {E}_{k}(\xi_k^*)\leq {E}_{k}(\xi_k), \forall k, {\rm ~with~} {{\pmb \xi}=(\xi_1, \ldots,\xi_K) \in \mathbb{P}}.
 \end{align}
 \end{theorem}
\begin{IEEEproof} We partition the feasible region \(\mathbb{P}\) into
\[
\mathbb{P}_{1a}=\left\{\pmb{\xi}:{\Gamma}_{1}\left(\pmb{\xi}\right)>\Delta_1 ~{\rm and }~{\Gamma}_{j}\left(\pmb{\xi}\right)\geq\Delta_j, ~j\neq 1\right\},
\]
and
 \[
 \mathbb{P}_{1b}=\left\{\pmb{\xi}:{\Gamma}_{1}\left(\pmb{\xi}\right)=\Delta_1 ~{\rm and }~{\Gamma}_{j}\left(\pmb{\xi}\right)\geq\Delta_j, ~j\neq 1\right\}.
 \]
For any \(\bar{\pmb{\xi}}\in \mathbb{P}_{1a}\), there exists a sufficiently small positive value \(\nu\) and \(\tilde{\pmb{\xi}}=\left(\bar{\xi}_1-\nu,\bar{\xi}_2,\cdots,\bar{\xi}_K\right) \in \mathbb{P}_{1b}\) such that \(\sum_{k=1}^{K}  {E}_{k}(\tilde{\xi}_k)\leq \sum_{k=1}^{K}  {E}_{k}(\bar{\xi}_k)\).
 Hence, the optimal solution \(\pmb{\xi}^*\in \mathbb{P}_{1b} \). Using similar arguments we can show that \( \pmb{\xi}^* \in \mathbb{P}_{kb}\) for any \(k\), where
 \[
 \mathbb{P}_{kb}=\left\{\pmb{\xi}:{\Gamma}_{k}\left(\pmb{\xi}\right)=\Delta_1 ~{\rm and }~{\Gamma}_{j}\left(\pmb{\xi}\right)\geq\Delta_j, ~j\neq k\right\}.
 \]
 Thus, \( \pmb{\xi}^* \in \mathbb{P}_{1b}\cap\cdots \cap\mathbb{P}_{Kb}\), i.e.,
 \begin{align}\label{Uplink_1}
     {\Gamma}_{k}\left(\pmb{\xi}^*\right)=\Delta_k, k=1,\cdots,K.
 \end{align} 
 
 (\ref{Uplink_1}) can be rewritten as 
 \[{\pmb W}\pmb{\xi}^*={\pmb b}.\]

Next, we prove (\ref{Th1}). Since \({E}_{k}(\xi_k)\) is a linear function of \(\xi_k\), (\ref{Th1}) is equivalent to  
\begin{align}
\xi_k^*\leq \xi_k, \forall k, {\rm ~with~} {{\pmb \xi}\in \mathbb{P}}.
 \end{align}
 We show this by contradiction. Otherwise assume that there exists
\begin{align*}
&\pmb{\xi}^{'}=(\xi_1^{'},\cdots,\xi_K^{'})=\left(c_1\xi_1^{*},\cdots,c_K\xi_K^{*}\right)
 \end{align*}
with some elements \({\xi}_k^{'} =c_k{\xi}_k^{*} <{\xi}_k^{*} \) in \(\mathbb{P}\). Without loss of generality, we assume \(c_1<1\) and \(c_k>0,~k\neq 1\). By (\ref{Uplink_1}), we have
\[
\frac{\mathcal{D}_{1}{\xi}_1^{*}}{\mathcal{U}_{1}{\xi}_1^{*}+\sum_{j\neq 1}\mathcal{I}_{1j}\xi_j^{*}+\mathcal{N}_{1}
	}=\Delta_1.
\]
In order to satisfy
\begin{align}
   {\Gamma}_{1}(\pmb{\xi}^{'})&=\frac{\mathcal{D}_{1}c_1{\xi}_1^{*}}{\mathcal{U}_{1}c_1{\xi}_1^{*}+\sum_{j\neq 1}\mathcal{I}_{1j}\xi_j^{'}+\mathcal{N}_{1}
	}\cr
	&\geq \Delta_1=\frac{\mathcal{D}_{1}{\xi}_1^{*}}{\mathcal{U}_{1}{\xi}_1^{*}+\sum_{j\neq 1}\mathcal{I}_{1j}\xi_j^{*}+\mathcal{N}_{1}
	}, 
\end{align}
 we should have  \(\sum_{j\neq 1}\mathcal{I}_{1j}\xi_j^{'}< c_1\sum_{j\neq 1}\mathcal{I}_{1j}\xi_j^{*}\), which implies at least one  \(c_k<c_1, k=2,\cdots,K\). Without loss of generality, we assume \(c_2<c_1\). Using similar arguments, one can show that at least one  \(c_k<c_2<c_1, k=2,\cdots,K\) to satisfy \({\Gamma}_{2}(\pmb{\xi}^{'})\geq\Delta_2\). Continuing in this way to satisfying \({\Gamma}_{k}(\pmb{\xi}^{'})\geq \Delta_k, k=1,\cdots,K-1\), we conclude that
\begin{align}\label{Uplink_3}
c_1>c_2>\cdots,>c_K>0.
\end{align}
Thus, we have 
\begin{align}
   {\Gamma}_{K}(\pmb{\xi}^{'})&=\frac{\mathcal{D}_{K}c_K{\xi}_K^{*}}{\mathcal{U}_{K}c_K{\xi}_K^{*}+\sum_{j\neq 1}\mathcal{I}_{Kj}c_j\xi_j^{*}+\mathcal{N}_{K} 
	} \cr
	&<{\Gamma}_{K}(\pmb{\xi}^{*})=\Delta_K,
\end{align}
which implies \(\pmb{\xi}^{'}\notin \mathbb{P}\) contradicting to our assumption. Then, we conclude the proof.
\end{IEEEproof}

To understand Theorem 3, consider the case \(K=2\). Then, the feasible region \(\mathbb P\) is
\begin{align*}
\xi_1\geq \frac{\Delta_1\mathcal{I}_{12}}{\mathcal{D}_{1}-\Delta_1\mathcal{U}_{1}} \xi_2+   \frac{\Delta_1}{\mathcal{D}_{1}-\Delta_1\mathcal{U}_{1}}\mathcal{N}_{1},
\end{align*}
and
\begin{align*}
\xi_2\geq \frac{\Delta_2\mathcal{I}_{21}}{\mathcal{D}_{2}-\Delta_2\mathcal{U}_{2}} \xi_1+   \frac{\Delta_2}{\mathcal{D}_{2}-\Delta_2\mathcal{U}_{2}}\mathcal{N}_{2}.
\end{align*}
Thus, \(\mathbb P\) is the shaded area as shown in Figure 3. It is straightforward to see that  \(\pmb{\xi}^{*}\) is the optimal solution which can simultaneously minimize the energy consumption of both sensors, since the feasible region is a cone.

\begin{figure}
\centering \scalebox{0.8}{\includegraphics[width=\columnwidth]{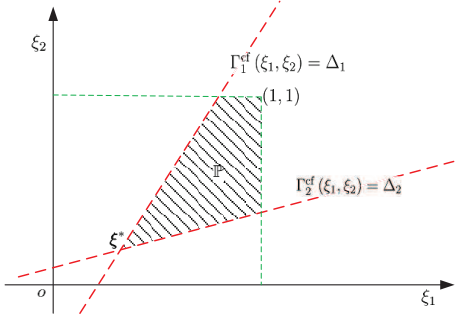}}
\centering \caption{The feasible region of P1 (shadow area).}
\end{figure}

	\subsection{Closed-form Asymptotically Optimal Solution to {\bf P2}}

 According to (\ref{joint_1}), (\ref{equvalents1}), and (\ref{joint_2}), {\bf P2} can be rewritten as
\begin{align}
	\mathop {\min }\limits_{\pmb{\mu}} &~ Q (1-\frac{\tau}{T})\rho_dN\sum_{l=1}^L\sum_{k=1}^K {\mu}_{l,k} \label{Down_opt.1}\\
		{s.t}. & ~ \left[N\left(\sum_{l=1}^{L}\sqrt{{\mu}_{l,k}\gamma_{l,k}}\right)^2+\sum_{l=1}^{L}{{\mu}_{l,k}\gamma_{l,k}}\right]\geq \mathcal{C}_k,~\forall k, \label{Down_opt.1.1}\\
		& ~  \mu_{l,k} \geq 0,~\forall l,k, \nonumber
		\end{align}
where \[\mathcal{C}_{k}=\frac{{E}_{k}(\xi_k)+E_0}{ (1-\frac{\tau}{T})B\zeta \rho_d N}.\]

Then {\bf P2} is non-convex due to the non-linear constraints in (\ref{Down_opt.1.1}). Since \(\sum_{l=1}^{L}{{\mu}_{l,k}\gamma_{l,k}}\leq \left(\sum_{l=1}^{L}\sqrt{{\mu}_{l,k}\gamma_{l,k}}\right)^2\), we drop the term \(\sum_{l=1}^{L}{{\mu}_{l,k}\gamma_{l,k}}\) in (\ref{Down_opt.1.1}), to obtain a relaxed problem \({\bf P2}^{'}\). Note that for massive MIMO, i.e., when \(N\) is large, \({\bf P2}^{'}\) well approximates \({\bf P2}\). It is not difficult to prove that the optimal solution to \({\bf P2}^{'}\) is obtained only when 
\begin{align}
   N\left(\sum_{l=1}^{L}\sqrt{{\mu}_{l,k}\gamma_{l,k}}\right)^2= \mathcal{C}_k,~\forall k.
\end{align}
 Let \(\vartheta_{l,k}=\sqrt{\mu_{l,k}}\), then \({\bf P2}^{'}\) becomes
\begin{align}
	\mathop {\min }\limits_{\pmb{\vartheta}} &~ Q (1-\frac{\tau}{T})\rho_dN\sum_{l=1}^L\sum_{k=1}^K {\vartheta}^2_{l,k} \label{Down_opt.2}\\
		{s.t}. & ~ \sum_{l=1}^{L}\sqrt{\gamma_{l,k}}{\vartheta}_{l,k}= \sqrt{\mathcal{C}_k/N},~\forall k,\nonumber\\
		& ~  \vartheta_{l,k} \geq 0,~\forall l,k. \nonumber
		\end{align}
It is easily seen that \({\bf P2}^{'}\) can be decomposed into the following \(K\) independent minimization problems for \(k=1,\cdots,K\):
\begin{align}
	\mathop {\min }\limits_{\pmb{\vartheta}_k} &~ Q (1-\frac{\tau}{T})\rho_dN\sum_{l=1}^L {\vartheta}^2_{l,k} \label{Down_opt.3}\\
		{s.t}. & ~ \sum_{l=1}^{L}\sqrt{\gamma_{l,k}}{\vartheta}_{l,k}= \sqrt{\mathcal{C}_k/N},\nonumber\\
		& ~  \vartheta_{l,k} \geq 0,~l=1,\cdots, L. \nonumber
		\end{align}
Using the method of the Lagrange multipliers, the closed-form optimal solution to (\ref{Down_opt.3}) is 
\begin{align}
    \vartheta_{l,k}^{*}=\frac{\sqrt{\gamma_{l,k}\mathcal{C}_k/N}}{\sum_{l=1}^{L}\gamma_{l,k}}, ~~l=1,\cdots, L.
\end{align}
It is noted that the optimal solution to (\ref{Down_opt.2}) is a feasible solution to (\ref{Down_opt.1}), and approaching the optimal solution to (\ref{Down_opt.1}) as \(N\) grows large. After finding \(\pmb{\vartheta}^{*}\),we get \(\mu_{l,k}^*={(\vartheta_{l,k}^*)}^2\). Further, we use \(\pmb{\mu}^{*}=\lambda\tilde{\pmb{\eta}}\) to find \(\lambda\) and \(\tilde{\pmb{\eta}}\). To guarantee the power constraints
\[
\sum_{k=1}^K \tilde{\eta}_{l,k} \leq 1, l=1,\cdots, L,
\]
we have
\[
\sum_{k=1}^K\mu_{l,k}^* \leq \lambda, l=1,\cdots, L.
\]
Thus, the minimum charging duration is 
\begin{align}
  \lambda^{*}=\mathop {\max }_{l=1,\ldots,L} \quad \sum_{k=1}^K \mu_{l,k}^*.
\end{align}
Next we find
\[\tilde{\eta}^{*}_{l,k}=\frac{\mu_{l,k}^*}{\lambda^{*}}.\]

	\section{Numerical results}
In this section, simulation results are provided to corroborate our theoretical analysis and to illustrate the gain due to our proposed system optimization. We consider a large square hall of \(50\times 50\) \({\rm m}^2\) with wrapped-around to avoid boundary effects. \(L=144\) APs are placed on the ceiling to form a square array with \(12\) APs in each column and row. \(K=20\) active sensors are randomly distributed in this area. The pilot sequences \(\pmb{\psi}_k,~k=1,\cdots, K,\) are randomly generated and fixed for all simulations. The parameters of the channel model is set according to \cite{S. Rao}. The large scale fading coefficient \(\beta_{l,k}\) is modeled as
\[
\beta_{l,k}=\mathcal{L}_{l,k}10^{\frac{{\sigma_{sh}z_{l,k}}}{10}}
\]
where \(\mathcal{L}_{l,k} (\rm dB)\) is the path loss and  \(10^{\frac{{\sigma_{\rm sh}z_{lk}}}{10}}\) is the shadow fading with standard deviation \(\sigma_{sh}=8\)dB and \(z_{l,k}\sim \mathcal {CN}(0,1)\). Similar to \cite{H. Q. Ngo_a}, we use the three-slope path-loss model .
{\small\begin{align}\label{eq:ploss}
&\mathcal{L}_{l,k}
=& \left\{
\begin{array}{l}
  -\mathcal{L}_0 - 35\log_{10} (d_{lk}), ~ \text{if} ~ d_{lk}>d_1,\\
  -\mathcal{L}_0 - 15\log_{10} (d_1) - 20\log_{10} (d_{lk}), ~ \text{if} ~ d_0< \!d_{lk}\leq d_1,\\
  -\mathcal{L}_0 - 15\log_{10} (d_1) - 20\log_{10} (d_{0}), ~ \text{if} ~ d_{lk} \leq d_0,\\
\end{array}%
\right.
 \end{align}}
with \(d_0=10\)m, \(d_1=50\)m, and
\begin{align}\label{eq:ploss1}
\mathcal{L}_0&\triangleq 46.3+33.9\log_{10}(f)-13.82\log_{10}(h_{\text{AP}})\nonumber\\
&-
(1.1\log_{10}(f)-0.7)h_{\text{s}}+(1.56\log_{10}(f)-0.8),
\end{align}
where $f=1900$MHz is the carrier frequency, $h_{\text{AP}}=7$m and $h_{\text{s}}=1.65$m denote the antenna height of APs and sensors, respectively. The transmit power is normalized by the noise power, which is given by
\[
\sigma^2 = {B}\times k_B\times T_0\times \kappa,
\]
where \(k_B= 1.381\times 10^{-23}{\rm J/K}\) is the Boltzmann constant and \(B\) is the bandwidth. \(T_0=290K\) and \(\kappa=9\)dB denote the noise temperature and the noise figure, respectively. 

To evaluate the spectrum efficiency, we use the per user throughput defined as
\begin{align}\label{Per user throughput}
    R_k=\frac{1-\tau/T}{2(1+\lambda)}B\log_2(1+\Gamma_k) ~~\mbox{bits/s}.
\end{align}
To account for the energy consumption due to pilots and circuits, \(E_0\) in (\ref{joint_1}) is set as
\[ E_0= (1+\lambda_0)\frac{\tau}{T}Q\rho_p+(1+\lambda_0)Q \rho_0, \]
where \(\rho_0=0.1\) mW is the ideal power consumption of each sensor, and \(\lambda_0=50\) is the maximum WPT duration allowed to guarantee the spectrum efficiency. In all examples, we choose the system parameters listed in
Table I.
\begin{table}[t!]
\renewcommand{\arraystretch}{1.3}
\caption{Simulation Parameters}
\label{table_parameters} \vskip-2mm
\centering
\begin{tabular}{|c|c|c|}
\hline
\bfseries parameter & \bfseries Meaning &\bfseries Value\\
\hline $L$ & Number of APs & 144\\
\hline $N$ & Number of antennas of each AP & 10\\
\hline $K$ & Number of active sensors & 20\\
\hline $B$ & Bandwidth & 20 MHz\\
\hline $T_c$ & Coherence time& 0.2 s\\
\hline $T$ & Number of symbols in each $T_c$ & 200\\
\hline $\tau$ & Length of pilot & 60\\
\hline $P_p$ & Pilot transmit power & 0.2mW \\
\hline $\rho_p$ & Normalized $P_p$ & $\rho_p=\frac{P_p}{\sigma^2}$ \\
\hline $P_u$ & Maximum uplink transmit power  & 20 mW \\
\hline $\rho_u$ & Normalized $P_u$  & $\rho_u=\frac{P_u}{\sigma^2}$ \\
\hline $\zeta$ & Energy conversion efficiency  & 1 \\
\hline $P_d$ & Maximum downlink transmit power  & 30 W \\
\hline $\rho_d$ & Normalized $P_d$  & $\rho_d=\frac{P_d}{\sigma^2}$ \\
\hline $\xi_k$ & Uplink power control coefficients & Optimized \\
\hline $\eta_{l,k}$ & Downlink power control coefficients  & Optimized \\
\hline $\lambda$ & WPT time duration  & Optimized \\
\hline
\end{tabular}
\end{table}
In addition, We fixed the time of data transmission in each period is one second, which implies that \(Q=1/T_c=5\). 

We first verify the accuracy of the closed-form expressions \(\tilde{\mathcal E}_{k}\) in (\ref{Per_B.2}) and \(\Gamma_k\) in (\ref{Per_B.3}) for cell-free IoT systems for one realization of large-scale fading \(\{\beta_{l,k}\}\). In Figure 4, the lower bounds \(\tilde{\mathcal E}_{k}, k=1,\cdots,K\) in (\ref{Per_B.2}), are compared with the simulation results obtained by (\ref{equ:sys5}) using 500 small-scale fading channel realizations, under the uniform power control, i.e., \(\tilde{\eta}_{l,k}=1/K, \forall l,k\). It is seen that the gap between the lower-bound \(\tilde{\mathcal E}_{k}\) and the simulation result is less than 10\%. This is because \(\mathbb{E}[|\mathcal{S}_{k1}|^2] \gg \mathbb{E}[|\mathcal{S}_{k2}+\mathcal{S}_{k3}|^2] \) in (17) as \(N\) is large. In Figure 5, the closed-form \(\Gamma_k\) in (\ref{Per_B.3}) is compared with the simulation results obtained by (\ref{equ:sys8}) using 500 small-scale fading channel realizations with full transmit power, i.e. \(\xi_k=1,\forall k\). It is seen that the closed-form expressions match well with the simulation results.	
	\begin{figure}
	\centering \scalebox{1}{\includegraphics[width=\columnwidth]{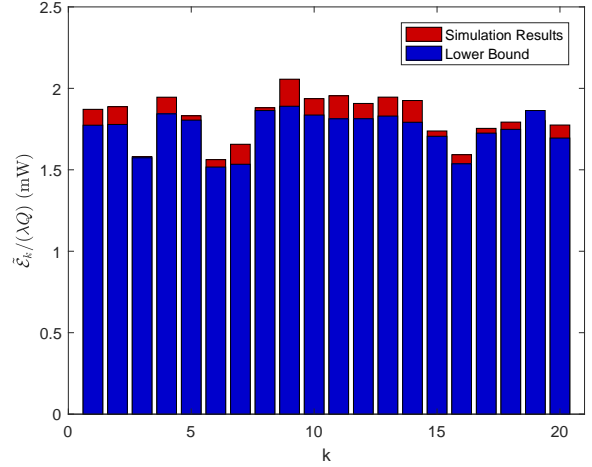}}
	\centering \caption{Tightness of the lower bound $\tilde{\mathcal E}_{k}$ in (\ref{Per_B.2}).}
\end{figure}

	\begin{figure}
	\centering \scalebox{1}{\includegraphics[width=\columnwidth]{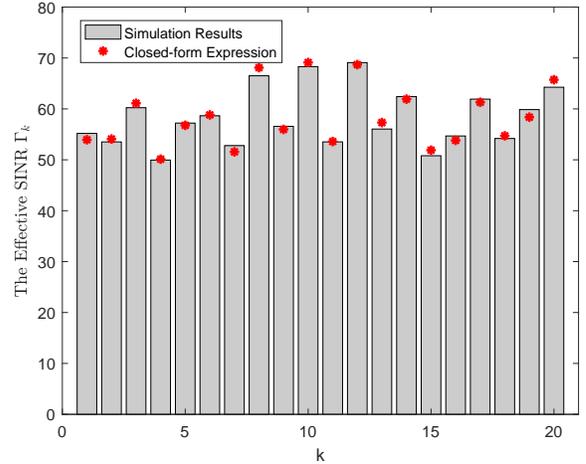}}
	\centering \caption{Accuratcy of the SINR expression in (\ref{Per_B.3}).}
\end{figure}

Next, we compare the uplink and downlink performances of three systems which are cell-free IoT, collocated massive MIMO, and small-cell IoT for 200 realizations of large-scale fading \(\{\beta_{l,k}\}\). Figure 6 shows the cumulative distribution function (CDF) of the amount of energy harvested per second, i.e., \(\tilde{\mathcal{E}}_k/(\lambda Q)\), for three systems. For cell-free IoT and collocated massive MIMO systems, the uniform power control scheme is adopted. For small-cell IoT, the \(k\)-th sensor is powered by its associated AP, i.e., \(\tilde{\eta}_{l,k}=1\) if \(\delta_{l,k}=1\). It can be seen that, the harvested energy of small-cell IoT is smaller than that of the other two systems due to the lower array gain. For collocated massive MIMO, the amount of harvested energy of the cell-boundary sensors is typically small, while that of the sensors adjacent to the AP is very high. Compared with the collocated massive MIMO, the distribution of the harvested energy in cell-free IoT is more concentrated, which result in the substantial improvement of the \(95\%\) likely performance. From Figure 6, it can be seen that the \(95\%\) likely performance of cell-free IoT is about 5 times higher than the collocated massive MIMO. Figure 7 plots the CDF of the effective SINR for three scenarios with full transmit power, i.e., \(\xi_k=1, \forall k\). Similarly as the amount of energy harvested, the  distribution of effective SINR is more concentrated, and the 95\% likely performance is significantly higher than that of the collocated massive MIMO and small-cell IoT.
	\begin{figure}
	\centering \scalebox{1}{\includegraphics[width=\columnwidth]{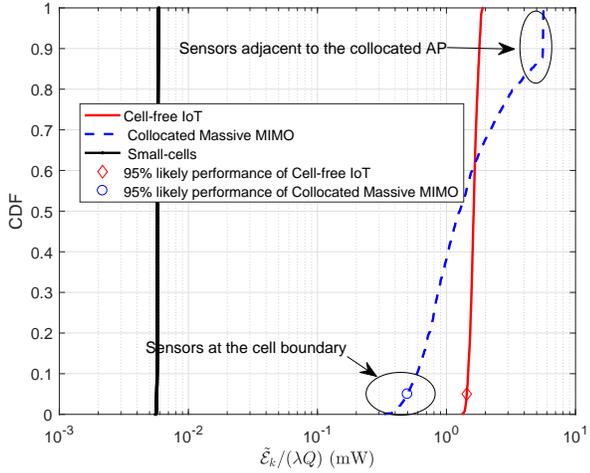}}
	\centering \caption{Downlink performance comparison in terms of $\tilde{\mathcal E}_{k}/(\lambda Q)$ (mW).}
\end{figure}
	\begin{figure}
	\centering \scalebox{1}{\includegraphics[width=\columnwidth]{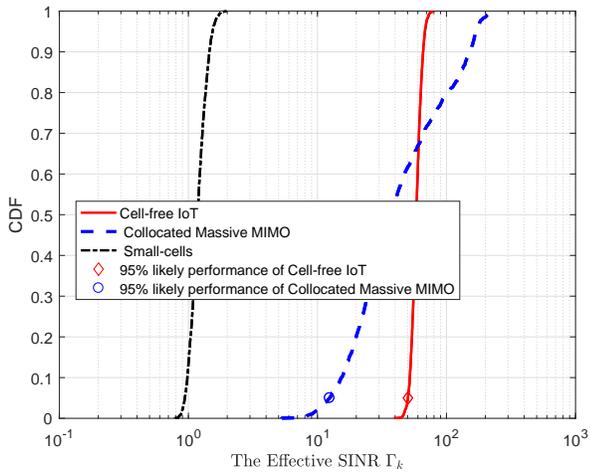}}
	\centering \caption{Uplink performance comparison in terms of SINR.}
\end{figure}

Finally, the performance of our joint downlink and uplink power control method is investigated. For comparison, we take the no power control scheme with \(\xi_k=1, \forall k\) and \(\tilde{\eta}_{l,k}=1/K\) as the benchmark. The result is taken over 200 realizations of large-scale fading \(\{\beta_{l,k}\}\). Figure 8 shows the total energy consumption \(\Xi_{\rm tr}\) given in (\ref{joint_2}) to support data transmission with given target SINR \(\Delta_k=30, \forall k,\). It can be seen that, the total energy consumption \(\Xi_{\rm tr}\) can be reduced by about 30\% using the joint downlink and uplink power control. On one hand, the energy consumption of each sensor can be reduced greatly to support the given target SINR using the uplink power control. On the other hand, the total energy consumption can be further reduced through the downlink power control. The CDF of the per user throughput is plotted in Figure 9. It can be seen that the per user throughput can be improved by 100\%, compared with the benchmark. In a word, the energy efficiency can be greatly improved through our joint power control method, in terms of both per user throughput and energy consumption.

	\begin{figure}
	\centering \scalebox{1}{\includegraphics[width=\columnwidth]{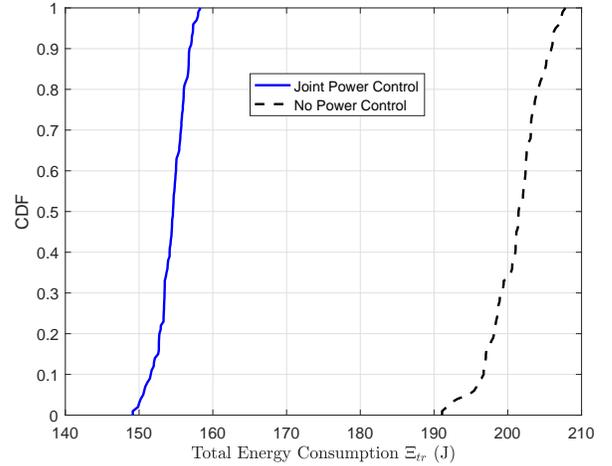}}
	\centering \caption{Performance gain due to joint power control in terms of total energy consumption \(\Xi_{\rm tr}\) to support data transmission with given target SINR \(\Delta_k=30, k=1,\cdots, K\) in each period.}
\end{figure}
		\begin{figure}
	\centering \scalebox{1}{\includegraphics[width=\columnwidth]{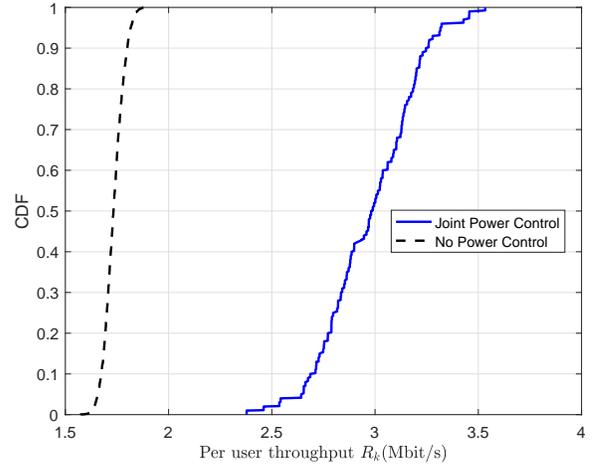}}
	\centering \caption{Performance gain due to joint power control in terms of per user throughput with \(\Delta_k=30,\forall k\).}
\end{figure}

	\section{Conclusions}
In this paper, we have propose a wirelessly powered cell-free IoT system and obtained the closed-form expressions of the downlink and uplink performance metrics, i.e., the amount of harvested energy for downlink, and the SINR for uplink. To minimize the total transmit power consumption
under the given SINR constraints, we proposed the joint downlink and uplink power control and provided closed-form solutions. Numerical results indicate that the proposed cell-free massive IoT system significantly outperforms its collocated massive MIMO and small-cell counterpart in terms of both downlink and uplink performances. And the proposed joint power control further boost the system performance.
	\vspace{0.2in}
	\section*{Appendix}
\subsection{Proof of Lemma 1}
\begin{IEEEproof}
Using (\ref{Per_A.6}),  the element of the correlation  matrix \({\rm cov}\left[\hat{\pmb g}_{l,k},~\hat{\pmb g}_{m,k}\right]\) in \(n\)-th row and \(\bar{n}\)-th column is
\[\begin{split}
&{\rm cov}\left[\hat{g}_{(l,n),k},~\hat{g}_{(m,\bar{n}),k}\right]\\
&=\mathbb{E}\left[\hat{g}_{(l,n),k}\hat{g}^H_{(m,\bar{n}),k}\right]-\mathbb{E}\left[\hat{g}_{(l,n),k}\right]\mathbb{E}\left[\hat{g}^H_{(m,\bar{n}),k}\right]\\
&=\mathbb{E}\left[\tau\rho_p\sum\nolimits_{i=1}^{K}\sum\nolimits_{j=1}^{K}{g}_{(l,n),i}{g}_{(m,\bar{n}),j}\pmb{a}_{l,k}^{H}\pmb{\varphi}_i\pmb{\varphi}_j^H\pmb{a}_{m,k}\right.\\
&\left.+\pmb{a}_{l,k}^{H}\pmb{w}_{(l,n)}\pmb{w}^H_{(m,\bar{n})} \pmb{a}_{m,k}
+\sqrt{\tau\rho_p}\sum\limits_{i=1}^{K}{g}_{(l,n),i}\pmb{a}_{l,k}^{H}\pmb{\varphi}_i\pmb{w}_{(m,{\bar n})}^H\pmb{a}_{m,k}\right.\\&\left.+\sqrt{\tau\rho_p}\sum\nolimits_{j=1}^{K}{g}_{(m,\bar{n}),j}\pmb{a}_{m,k}^{H}\pmb{\varphi}_j\pmb{w}_{(l,n)}^H\pmb{a}_{l,k}\right]\\
&\overset{(a)}{=}\mathbb{E}\left[\pmb{a}_{l,k}^{H}\pmb{w}_{(l,n)}\pmb{w}^H_{(m,\bar{n})} \pmb{a}_{m,k}\right]\\
&={\rm tr}\left[\mathbb{E}\left(\pmb{w}_{(l,n)}\pmb{w}^H_{(m,\bar{n})} \pmb{a}_{m,k}\pmb{a}_{l,k}^{H}\right)\right]\\
&\overset{(b)}{=}0
\end{split}
\]	
where (a) is obtained according the independence of \({g}_{(l,n),i}\) and \({g}_{(m,\bar{n}),j}\), while (b) is obtained according to the independence of \(\pmb{w}_{(l,n)}\) and \(\pmb{w}_{(m,\bar{n})}\). Since each element is zero, the correlation matrix is zero matrix, i.e.,
\begin{align}\label{App_A.1}
{\rm cov}\left[\hat{\pmb g}_{l,k},~\hat{\pmb g}_{m,k}\right]={\pmb 0}.
\end{align}
Using (\ref{Per_A.6}), we can obtain

  \begin{align}
&\mathbb{E}\left[\hat{g}_{(l,n),k}^2,\hat{g}_{(m,\bar{n}),k}^2\right]\nonumber \\
&=\mathbb{E}\left[\left(\pmb{a}_{l,k}^{H}\left[\sqrt{\tau\rho_p}\sum\nolimits_{i=1}^{K}{g}_{(l,n),i}\pmb{\varphi}_i+\pmb{w}_{(l,n)}\right]\right)^2\right. \nonumber\\
&\times\left.\left(\pmb{a}_{m,k}^{H}\left[\sqrt{\tau\rho_p}\sum\nolimits_{j=1}^{K}{g}_{(m,\bar{n}),j}\pmb{\varphi}_j+\pmb{w}_{(m,\bar{n})}\right]\right)^2\right]\nonumber\\
&=\mathbb{E}\left[(\tau\rho_p\sum\limits_{i=1}^{K}{g}^2_{(l,n),i}\pmb{a}_{l,k}^{H}\pmb{\varphi}_i\pmb{a}_{l,k}^{H}\pmb{\varphi}_{i}+\pmb{a}_{l,k}^{H}\pmb{w}_{(l,n)}\pmb{a}_{l,k}^{H}\pmb{w}_{(l,n)} )\right.\times \nonumber\\ &\left.(\tau\rho_p\sum\limits_{j=1}^{K}{g}^2_{(m,{\bar n}),j}\pmb{a}_{m,k}^{H}\pmb{\varphi}_j\pmb{a}_{m,k}^{H}\pmb{\varphi}_{j}+\pmb{a}_{m,k}^{H}\pmb{w}_{(m,{\bar n})}\pmb{a}_{m,k}^{H}\pmb{w}_{(m,{\bar n})} )\right]\nonumber\\
&=\mathbb{E}\left[\hat{g}^2_{(l,n),k}\right]\mathbb{E}\left[\hat{g}^2_{(m,\bar{n}),k}\right]\label{App_A.2}.
\end{align}

Thus, we have
\begin{align}\label{App_A.3}
&{\rm cov}\left[\hat{g}^2_{(l,n),k},~\hat{g}^2_{(m,\bar{n}),k}\right]\nonumber \\
&=\mathbb{E}\left[\hat{g}^2_{(l,n),k}\hat{g}^2_{(m,\bar{n}),k}\right]-\mathbb{E}\left[\hat{g}^2_{(l,n),k}\right]\mathbb{E}\left[\hat{g}^2_{(m,\bar{n}),k}\right]=0.
\end{align}

By definition of \(\ell_2\) norm, we have
\[\begin{split}
&\mathbb{E}\left[\left\|\hat{\pmb g}_{l,k}\right\|^2\left\|\hat{\pmb g}_{m,k}\right\|^2\right]\\
&=\mathbb{E}\left[\left(\sum\nolimits_{n=1}^N\hat{g}^2_{(l,n),k} \right)\left(\sum\nolimits_{{\bar n}=1}^N\hat{g}^2_{(m,\bar{n}),k} \right)\right]\\
&=\sum\nolimits_{n=1}^N\sum\nolimits_{{\bar n}=1}^N\mathbb{E} \left[\hat{g}^2_{(l,n),k}\hat{g}^2_{(m,\bar{n}),k}\right]\\
&\overset{(a)}{=}\sum\nolimits_{n=1}^N\sum\nolimits_{{\bar n}=1}^N\mathbb{E}\left[\hat{g}^2_{(l,n),k}\right]\mathbb{E}\left[\hat{g}^2_{(m,\bar{n}),k}\right]\\
&=\mathbb{E}\left[\sum\nolimits_{n=1}^N\hat{g}^2_{(l,n),k}\right]\mathbb{E}\left[\sum\nolimits_{{\bar n}=1}^N\hat{g}^2_{(m,\bar{n}),k}\right]\qquad \qquad\\
&=\mathbb{E}\left[\left\|\hat{\pmb g}_{l,k}\right\|^2\right]\mathbb{E}\left[\left\|\hat{\pmb g}_{m,k}\right\|^2\right],
\end{split}
\]
where (a) is obtained by (\ref{App_A.3}). Thus, we have
\begin{align}\label{App_A.4}
{\rm cov}\left[\left\|\hat{\pmb g}_{l,k}\right\|^2,~\left\|\hat{\pmb g}_{m,k}\right\|^2\right]=0.\qquad\qquad
\end{align}

\end{IEEEproof}	

\subsection{Proof of Theorem 1}

\begin{IEEEproof}		
First, we compute the power of \(\mathcal{A}_1\). Since \(\hat{\pmb{g}}_{l,k}\) and \(\tilde{\pmb{g}}_{l,k}\) are independent, we have
\begin{align}\label{App_B.1}
		\left|\mathcal{A}_1\right| ^2  &=\xi_k\rho_u\left|\sum\nolimits_{l=1}^L\mathbb{E}\left[\hat{\pmb{g}}_{l,k}^{H}(\hat{\pmb{g}}_{l,k}+\tilde{\pmb{g}}_{l,k}) \right]  \right| ^2\nonumber \\
		&=\xi_k\rho_uN^2\left|\sum\nolimits_{l=1}^L\gamma_{l,k} \right| ^2.\end{align}
		
Next, we compute the power of \(\mathcal{A}_2\). Since \(\hat{\pmb{g}}_{l,k}\) and \(\tilde{\pmb{g}}_{l,k}\) are independent, and \[\mathbb{E}\left[\left\|\hat{\pmb{g}}_{l,k}\right\|^4\right]=\mathbb{E}\left[\left\|\hat{\pmb{g}}_{l,k}\right\|^2\right]^2+\mathbb{D}\left[\left\|\hat{\pmb{g}}_{l,k}\right\|^2\right]=N(N+1)\gamma^2_{l,k},\]
the power of \(\mathcal{A}_2\) can be expressed as (\ref{App_B.2}) given at next page, where step (a) is obtained by using Lemma 1.
\begin{figure*}

\begin{align}\label{App_B.2}
    	&\mathbb{E}\left[\left|\mathcal{A}_2\right|^2  \right] =\rho\xi_k\sum\limits_{l=1}^{L}\sum\limits_{m=1}^{L}\mathbb{E}\left\{\left(\hat{\pmb{g}}_{l,k}^{H}\pmb{g}_{l,k}-\mathbb{E}\left[\hat{\pmb{g}}_{l,k}^{H}\pmb{g}_{l,k}\right]\right) \left(\hat{\pmb{g}}_{m,k}^{H}\pmb{g}_{m,k}-\mathbb{E}\left[\hat{\pmb{g}}_{m,k}^{H}\pmb{g}_{m,k}\right]\right) \right\}	\nonumber \\
    	& =\rho\xi_k\sum\limits_{n=1}^{N}\sum\limits_{m=1}^{L}\mathbb{E}\left\{\hat{\pmb{g}}_{l,k}^{H}\pmb{g}_{l,k}\hat{\pmb{g}}_{m,k}^{H}\pmb{g}_{m,k}-\hat{\pmb{g}}_{lk}^{H}\pmb{g}_{l,k}\mathbb{E}\left[\hat{\pmb{g}}_{m,k}^{H}\pmb{g}_{m,k}\right]-\mathbb{E}\left[\hat{\pmb{g}}_{l,k}^{H}\pmb{g}_{l,k}\right] \hat{\pmb{g}}_{m,k}^{H}\pmb{g}_{m,k}+\mathbb{E}\left[\hat{\pmb{g}}_{l,k}^{H}\pmb{g}_{l,k}\right] \mathbb{E}\left[\hat{\pmb{g}}_{m,k}^{H}\pmb{g}_{m,k}\right]
		\right\}  \nonumber\\
		 &=\rho\xi_k\sum\limits_{l=1}^{L}\sum\limits_{m=1}^{L}\mathbb{E}\left[\hat{\pmb{g}}_{l,k}^H\left(\hat{\pmb{g}}_{l,k}+\tilde{\pmb{g}}_{l,k}\right)\hat{\pmb{g}}_{m,k}^H\left(\hat{\pmb{g}}_{m,k}+\tilde{\pmb{g}}_{m,k}\right)\right]-\rho\xi_kN^2\sum_{l=1}^{L}\sum_{m=1}^{L}\gamma_{l,k}\gamma_{m,k}\nonumber\\
		 &=\rho_u\xi_k\sum\limits_{l=1}^{L}\left[\mathbb{E}\left[\left\|\hat{\pmb{g}}_{l,k}\right\|^4\right]+\mathbb{E}\left[\left|\hat{\pmb{g}}^H_{l,k}\tilde{\pmb{g}}_{l,k}\right|^2\right]+\sum\limits_{m\neq l}\mathbb{E}\left[\left\|\hat{\pmb{g}}_{l,k}\right\|^2\left\|\hat{\pmb{g}}_{m,k}\right\|^2\right]+\sum\limits_{m\neq l}\mathbb{E}\left[\hat{\pmb{g}}^H_{l,k}\tilde{\pmb{g}}_{l,k}\hat{\pmb{g}}^H_{m,k}\tilde{\pmb{g}}_{m,k}\right]\right.\nonumber\\
		 &\left.+\sum\limits_{m=1}^{L}\left\{\mathbb{E}\left[\left\|\hat{\pmb{g}}_{l,k}\right\|^2\hat{\pmb{g}}_{m,k}^H\tilde{\pmb{g}}_{m,k}\right]+\mathbb{E}\left[\left\|\hat{\pmb{g}}_{m,k}\right\|^2\hat{\pmb{g}}_{l,k}^H\tilde{\pmb{g}}_{l,k}\right]\right\}\right]-\rho\xi_kN^2\sum\limits_{l=1}^{L}\sum\limits_{m=1}^{L}\gamma_{l,k}\gamma_{m,k}\nonumber\\
		&\overset{(a)}{=}\rho\xi_k\sum\limits_{l=1}^{L}\left[ N(N+1)\gamma_{lk}^2+N\gamma_{lk}\left(\beta_{lk}-\gamma_{lk}\right)+N^2\sum\limits_{m\neq l}\gamma_{lk}\gamma_{mk}\right]-\rho\xi_kN^2\sum\limits_{l=1}^{L}\sum\limits_{m=1}^{L}\gamma_{lk}\gamma_{mk}\nonumber\\
		&=\rho\xi_kN\sum_{l=1}^{L}\gamma_{lk}\beta_{lk}.
\end{align}
\end{figure*}

	 Then, the power of \(\mathcal{A}_3\) can be expressed as
   \begin{align}\label{App_B.3}
		\mathbb{E}\left[\left|\mathcal{A}_3\right|^2 \right] =\sum\nolimits_{j \neq k}\rho_u\xi_j\mathbb{E}\left[\left|\sum\nolimits_{l=1}^L\hat{\pmb{g}}_{l,k}^{H}\pmb{g}_{l,j}\right|^2\right],
        \end{align}
        where \(\mathbb{E}\left[|\sum_{l=1}^L\hat{\pmb{g}}_{l,k}^{H}\pmb{g}_{l,j}|^2\right]\) can be calculated as (\ref{App_B.4}) shown at next page, with step (a) is obtained by the equation (\ref{App_B.5}) shown at next page. Substituting (\ref{App_B.4}) into (\ref{App_B.3}), we obtain (\ref{App_B.6}) shown at next page.
 
\begin{figure*}
	\begin{flalign}\label{App_B.4}      &\mathbb{E}\left[\left|\sum\limits_{l=1}^L\hat{\pmb{g}}_{l,k}^{H}\pmb{g}_{l,j}\right|^2\right]=\mathbb{E}\left[\left|\sum\limits_{l=1}^L\sum\limits_{n=1}^N{g}_{(l,n),j}\hat{{g}}_{(l,n),k}^{*}\right|^2\right]\nonumber\\
&=\mathbb{E}\left[\left|\sum\limits_{l=1}^L\sum\limits_{n=1}^N{g}_{(l,n),j}\left(\pmb{a}_{l,k}^{H}\left[\sqrt{\tau\rho_p}\sum\limits_{i=1}^{K}{g}_{(l,n),i}\pmb{\varphi}_i+\pmb{w}_{(l,n)}\right]\right)^*
			\right|^2\right]\nonumber\\ 
			&=\mathbb{E}\left[\left|\sqrt{\tau\rho_p}\sum\limits_{l=1}^L\sum\limits_{n=1}^N{g}_{(l,n),j}\sum\limits_{i=1}^{K}{g}^{*}_{(l,n),i}\pmb{a}_{l,k}^T\pmb{\varphi}_i^{*}+\sum\limits_{l=1}^L\sum\limits_{n=1}^N{g}_{(l,n),j}\pmb{a}_{l,k}^{T}\pmb{w}_{(l,n)}^{*}\right|^2\right]\nonumber\\
		 &=\mathbb{E}\left[\left|\sum\limits_{l=1}^L\sum\limits_{n=1}^N{g}_{(l,n),j}\pmb{a}_{l,k}^T\pmb{w}_{(l,n)}^{*}\right|^2\right]+\tau\rho_p\mathbb{E}\left[\left|\sum\limits_{l=1}^L\sum\limits_{i=1}^{K}{\pmb g}^{H}_{l,i}{\pmb g}_{l,j}\pmb{a}_{l,k}^T\pmb{\varphi}_i^{*}\right|^2\right]\nonumber\\ 
			 &=\sum\limits_{l=1}^L\beta_{l,j}N\left\|\pmb{a}_{l,k}\right\|^2_2+\tau\rho_p\mathbb{E}\left[\sum\limits_{l=1}^L\sum\limits_{i=1}^{K}\sum\limits_{\bar{l}=1}^L\sum\limits_{\bar{i}=1}^{K}{\pmb g}^{H}_{l,i}{\pmb g}_{l,j}{\pmb g}^{H}_{\bar{l},\bar{i}}{\pmb g}_{\bar{l},j}\pmb{a}_{l,k}^T\pmb{\varphi}_i^{*}\pmb{a}_{{\bar l},k}^T\pmb{\varphi}_{\bar i}^{*}\right]\nonumber\\ 
				 &\overset{(a)}{=}\sum\limits_{l=1}^L\beta_{l,j}N\left\|\pmb{a}_{l,k}\right\|^2_2+\tau\rho_p\left(N^2\left|\sum\limits_{l=1}^L\beta_{l,j}\pmb{\psi}_j^H\pmb{a}_{l,k}\right|^2+\sum\limits_{l=1}^L\sum\limits_{i=1}^{K}N\beta_{l,j}\beta_{l,i}\left|\pmb{\psi}_i^H\pmb{a}_{l,k}\right|^2\right)\\
				 \nonumber \\
				 &\mathbb{E}\left[{\pmb g}^{H}_{l,i}{\pmb g}_{l,j}{\pmb g}^{H}_{\bar{l},\bar{i}}{\pmb g}_{\bar{l},j}\right]=\left\lbrace
\begin{aligned}
&\mathbb{E}\left[\left\|{\pmb g}_{l,j}\right\|^4 \right]= N(N+1)\beta_{l,j}^2 &\qquad \bar{l}=l, \bar{i}=i=j\\
&\mathbb{E}\left[\left| {\pmb g}_{l,j}^H{\pmb g}_{l,i}\right|^2\right] = N\beta_{l,j}\beta_{l,i} &\qquad \bar{l}=l, \bar{i}=i\neq j\\
&\mathbb{E}\left[\left\|{\pmb g}_{l,j}\right\|^2\left\|{\pmb g}_{\bar{l},j}\right\|^2 \right]= N^2\beta_{l,j} \beta_{{\bar l},j} &\qquad \bar{l}\neq l, \bar{i}=i=j\\
&0  &\qquad {\rm otherwise}\\
\end{aligned}
 \right. \label{App_B.5}\\
 \nonumber \\
 &\mathbb{E}\left[\left| \mathcal{A}_3\right| ^2 \right] = \sum_{j \neq k}\rho_uN\xi_j\left[\sum_{l=1}^L\beta_{l,j}\left\|\pmb{a}_{l,k}\right\|^2+\tau\rho_p\left(N\left|\sum_{l=1}^L\beta_{l,j}\pmb{\psi}_j^H\pmb{a}_{l,k}\right|^2+\sum_{l=1}^L\sum_{i=1}^{K}\beta_{l,j}\beta_{l,i}\left|\pmb{\psi}_i^H\pmb{a}_{l,k}\right|^2\right)\right] \quad \quad\quad\label{App_B.6}
			\end{flalign}
	\end{figure*}	

Finally, we compute the power of \(\mathcal{A}_4\). Due to the independence of \(\hat{\pmb{g}}_{l,k}\) and \(\tilde{\pmb{g}}_{l,k}\), we have
			\begin{align}\label{App_B.7}
			\mathbb{E}\left[\left| \mathcal{A}_4\right| ^2 \right] =\mathbb{E}\left[ \left|\sum_{l=1}^L\hat{\pmb{g}}_{l,k}^{H}\pmb{n}_l \right|^2\right]=N\sum_{l=1}^L\gamma_{l,k}.
			\end{align}
            Plugging (\ref{App_B.1}),(\ref{App_B.2}),(\ref{App_B.6}), and (\ref{App_B.7}) into (\ref{equ:sys8}), we obtain (\ref{Per_B.3}).
	\end{IEEEproof}

\end{document}